# Harmonics Mitigation of Industrial Power System Using Passive Filters

ZUBAIR AHMED MEMON*, MOHAMMAD ASLAM UQUAILI**, AND MUKHTIAR ALI UNAR***

## ABSTRACT

With the development of modern industrial technology a large number of non-linear loads are used in power system, which causes harmonic distortion in the power system. At the same time the power quality and safe operation becomes inferior. Therefore mitigation of harmonics is very necessary under the situation. This paper presents the design of two passive filters to reduce the current harmonics produced by nonlinear loads in industrial power system. Matlab /simlink software has been used for the simulation purpose. The results have been obtained with and without installation of filters and then it is observed that after installation of filters harmonics of the current are reduced and power factor is improved.

## 1.      INTRODUCTION

Nowadays industries prefer to use power electronics based devices due to their effectiveness. Though these power electronics based devices are advantageous to the electronics and electrical industry, these devices generate and inject the harmonics in the power industry. These harmonics are known as electrical disturbances which are the main cause of the power quality associated harms. The main problems due to the harmonics are additional power losses in the electrical equipment, irregular function of protective devices, errors in measurement of metering devices and interference with the telecommunication lines. Therefore mitigation of harmonics and improvement of the power quality is essential under the situation.

In the literature several studies have been presented regarding the harmonic mitigation by using different types of filters [1-6]. Passive filter is one of them and has been investigated for the harmonic mitigation. Low cost, simple design and high reliability are main advantages of passive filters [7].

In this paper effectiveness and design of single-tuned filter and second order high pass filter has been investigated for suppressing the harmonic currents in industrial power system.

## 2.      PASSIVE FILTERS

For mitigating the harmonic distortion passive filtering is the simplest conventional solution [8]. Passive elements

*       Assistant Professor, Department of Electrical Engineering, Mehran University of Engineering & Technology, Jamshoro.
**      Professor, Department of Electrical Engineering, Mehran University of Engineering & Technology, Jamshoro.
***     Professor, Department of Computer Systems Engineering, Mehran University of Engineering & Technology, Jamshoro.

like resistance, inductance and capacitance are used by the passive filters to control the harmonics. Common types of passive filters and their configurations are depicted in Fig. 1.

The shunt connection of passive filters with the power system provides least impedance path to the harmonic current at tuning frequency. As compared to the shunt filter series filter is designed to carry full load current therefore they need over current protection devices. Whereas shunt passive filter carries a fraction of series filter current. The series filter is relatively more expensive hence shunt passive filter is commonly used as harmonic filter. Furthermore it also provides reactive power at system operating frequency.

## 3.    SINGLE TUNED FILTER DESIGN

The most commonly used passive filter is the single-tuned filter. This filter is simple and least expensive as compared with other means for mitigating the harmonic problems [9-10]. The LC STF (Single Series Filter) is most common and inexpensive type of passive filter. This filter is connected in shunt with the main distribution system and is tuned to present low impedance to a particular harmonic frequency. Therefore, harmonic currents are diverted from the least impedance path through the filter. For designing the single-tuned filter it is essential to select the appropriate capacitor value that enables good power factor at system frequency. The circuit diagram of the STF is depicted in Fig. 1.

The impedance versus frequency curve of this filter is shown in Fig. 2.

In design of the filter, the proper selection of the capacitor size is very essential from power factor point of view. The relation between capacitor reactance and reactive power is expressed as:

$$Q_{Filter} = \frac{V_{cap}^2}{X_c} \tag{1}$$

where $V_{cap}^2$ is the line voltage of the capacitor in volts, Q is the reactive power in kVAR and $X_c$ is the capacitive reactance of capacitor in ohms.

The Equation (2) of the capacitive reactance is given by:

$$C = \frac{1}{2\pi f C} \tag{2}$$

From Equation (2) the value of capacitance (in farad) is calculated as:

$$C = \frac{1}{2\pi f X_C} \tag{3}$$

The resonance condition will occur when capacitive reactance is equal to inductive reactance as:

$$X_L = X_C \tag{4}$$

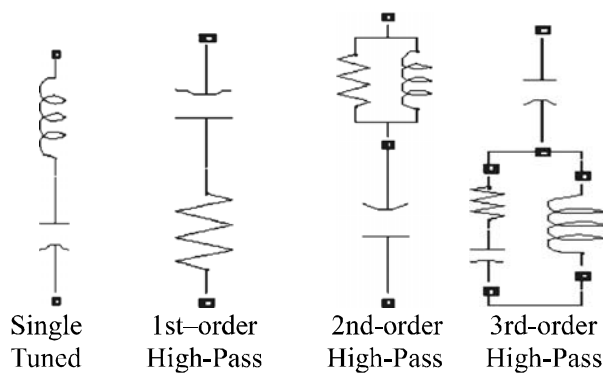

Single Tuned    1st–order High-Pass    2nd-order High-Pass    3rd-order High-Pass

*FIG. 1. PASSIVE POWER FILTERS CONFIGURATIONS*

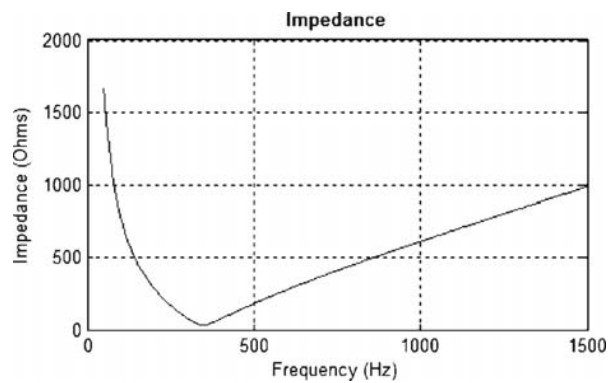

*FIG. 2.. THE CHARACTERISTIC OF THE SINGLE TUNED FILTER*

The Equation (4) can be rewritten as:

$$2\pi fL = \frac{1}{2\pi fC} \qquad (5)$$

The inductive value of the filter can be obtained from Equation (5) as:

$$L = \frac{1}{\left(2\pi f\right)^2 C} \qquad (6)$$

The resistance of filter depends on the quality factor (Q) by which sharpness of the tuning is measured. Mathematically quality factor is defined as:

$$Q = \frac{\sqrt{\dfrac{L}{C}}}{R} \qquad (7)$$

The resistive value of the filter can be obtained by selecting the quality factor in the range of 20<Q<100 [11].

The larger value of the quality factor gives the best reduction in harmonic reduction. However, it is necessary to take care of the harmonic frequencies because these harmonic current frequencies will also follow the least impedance path. These currents cause the increased power loss. Therefore it is necessary to perform the computer based harmonic simulation for analyzing the performance of the filters.

## 4. SECOND ORDER HIGH PASS FILTER

The second order high-pass filter is similar to single-tuned filter where L and R are connected in parallel instead of series as shown in Fig. 1. The second-order high pass filter provides good filtering performance and it decreases the energy losses at fundamental frequency.

The impedance of second-order high pass filter is given by Equation (8):

$$Z = \frac{1}{j\omega C} + \frac{1}{\left(\dfrac{1}{R} + \dfrac{1}{j\omega L}\right)} \qquad (8)$$

The corner frequency of the filter is achieved by Equation (5)

The quality factor of second-order high pass filter is different as compared to single tuned filter and it is reciprocal of the quality factor of the single tuned filter as given by the Equation (9):

$$Q = \frac{R}{X_C} = \frac{R}{X_L} \qquad (9)$$

The characteristic of second-order high pass filter is shown in Fig. 3.

For second order high pass filter the typical values of quality factor are between 0.5 and 5 [12].

At higher value of quality factor the filters give the superior filtering performance as compared to the filters with smaller value of quality factor.

The other points that are essential in selection of the quality factor are as under:

☐ Filter's tuning frequency

☐ Losses

☐ Concerns for telephone interference (if exists)

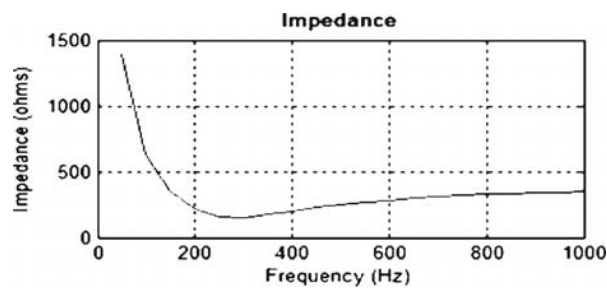

*FIG. 3. THE CHARACTERISTIC OF THE SECOND-ORDER HIGH PASS FILTER*

## 5.    SELECTION OF FILTERS

At lower harmonic frequencies the most of the waveforms have large percentage of harmonic distortion as compared to the high harmonic frequencies. For that reason single tuned filters are designed to suppress these lower harmonic frequencies. For suppressing the harmonics of six pulse ac to dc converter four single tuned filters are used for the 5th, 7th, 11th, and 13th harmonics and one second order high pass filter is used for eliminating the high order frequencies [13] as shown in Fig. 4.

## 6.    ANALYSIS OF SIMULATION RESULTS

In this paper, three-phase ac to dc converter has been simulated with and without proposed passive filters in the matlab/simulink environment.

The circuit parameters used in simulation are presented in Table 1.

Fig. 5 shows the three-phase supply currents of the system without passive filters. As it is clear that currents are distorted therefore these currents contain the harmonics.

Figs. 6-7 show the supply current without passive filters (only one phase current has been shown for the clearness) and its frequency spectrum, respectively. It is clear that the THD (Total Harmonic Distortion) of the current is 20.77%.

**TABLE 1. SIMULATION PARAMETERS [14]**

| | |
|---|---|
| Supply Voltage | Vs=220Vrms |
| Supply/Line Inductance | Ls=0.0016 Henery |
| Rectifier Front-End Inductance | LL=0.023 Henery |
| Capacitance of the Load | C=50 Micro Farad |
| Resistance of the Load | RL=78 Ohm |

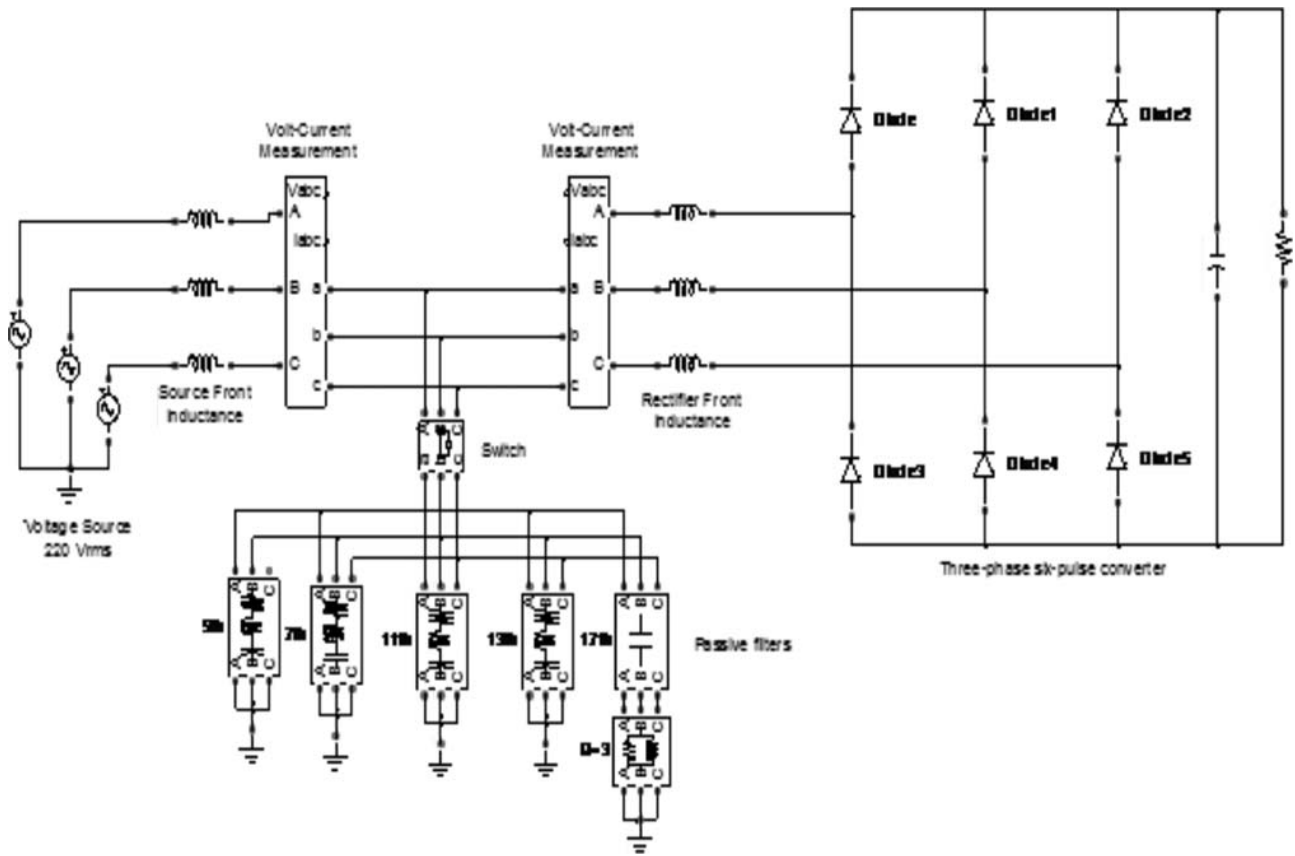

*FIG. 4. INDUSTRIAL POWER SYSTEM MODEL WITH PASSIVE FILTERS*

Without passive filters the total harmonic distortion of the current is above the range specified by the power quality standards.

To follow the recommended IEEE 519 power harmonic standards the total harmonic distortion must be less than 5%. This can be obtained by connecting the passive filters to the system. For reducing the THD below 5% passive filters have been designed. The parameters of the proposed passive filters have been shown in Table 2.

After connecting the filters the three-phase supply currents become sinusoidal and harmonics are decreased below 5%. Thus the source current becomes in phase with the supply voltage, the power factor of the source is then near to the unity. The simulation results are presented in Figs. 7-11 which show good filtering performance of the proposed filters.

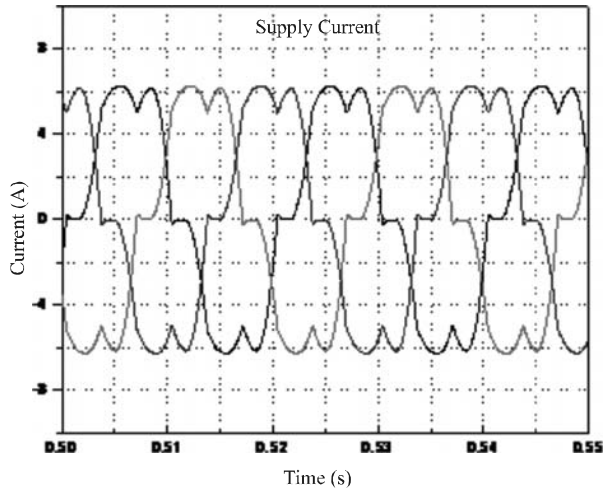

*FIG. 5. THREE-PHASE SUPPLY CURRENTS WITHOUT FILTERS.*

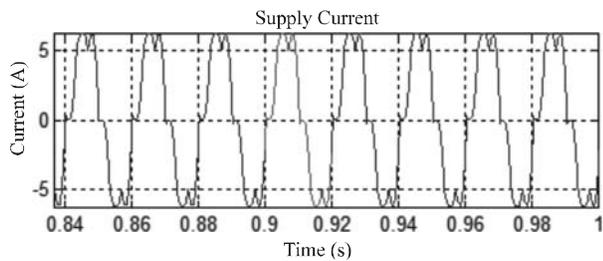

*FIG. 6. SUPPLY CURRENT WITHOUT PASSIVE FILTERS*

# 7. CONCLUSION

A PC-based design method of two common types of passive filters namely single tuned filter and second order high pass filter has been presented in this paper. The proposed filters reduce the total harmonic distortion of the source current at a high level of expectation from 20.77-4.32% in the simulation. Our results meet the IEEE 519 recommended harmonic standards.

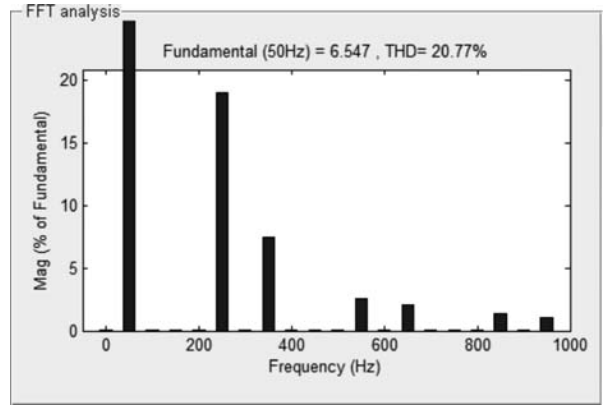

*FIG. 7. FREQUENCY SPECTRUM OF SUPPLY CURRENT WITHOUT PASSIVE FILTERS*

**TABLE 2. VALUES OF DESIGNED FILTERS**

| C (F) | L (H) | R (Ω) |
|---|---|---|
| $C_{5th}$ = 11.09e-6 | $L_{5th}$ = 0.0365 | $R_{5th}$ = 0.54 |
| $C_{7th}$ = 11.09e-6 | $L_{7th}$ = 0.0186 | $R_{7th}$ = 0.38 |
| $C_{11th}$ = 11.09e-6 | $L_{11th}$ = 0.0075 | $R_{11th}$ = 0.24 |
| $C_{13th}$ = 11.09e-6 | $L_{13th}$ = 0.0054 | $R_{13th}$ = 0.21 |
| $C_{HP}$ = 11.09e-6 | $L_{HP}$ = 0.0031 | $R_{HP}$ = 49.66 |

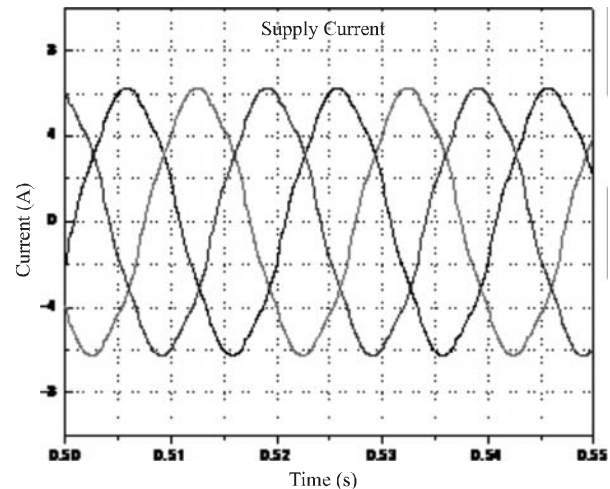

*FIG. 8. THREE PHASE SUPPLY CURRENTS WITH FILTERS*

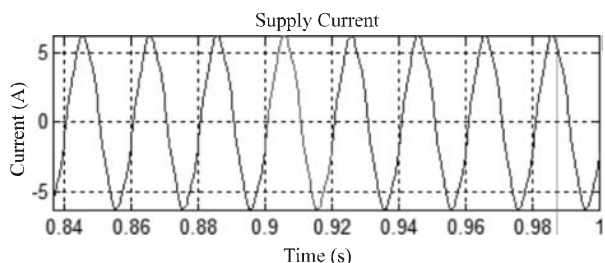

*FIG. 9. SUPPLY CURRENT WITH FILTERS*

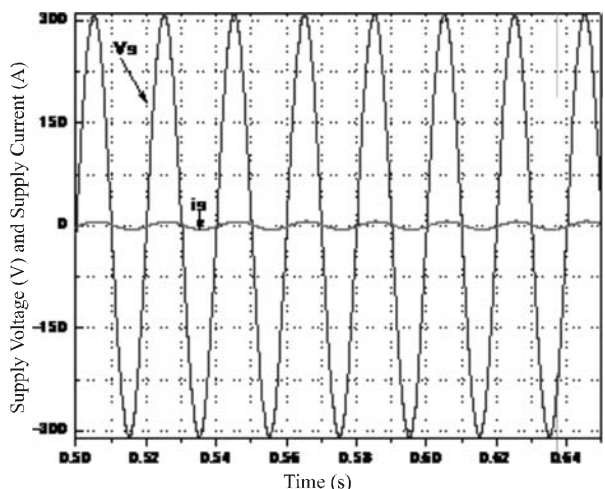

*FIG. 10. FREQUENCY SPECTRUM OF SUPPLY CURRENT WITH FILTERS.*

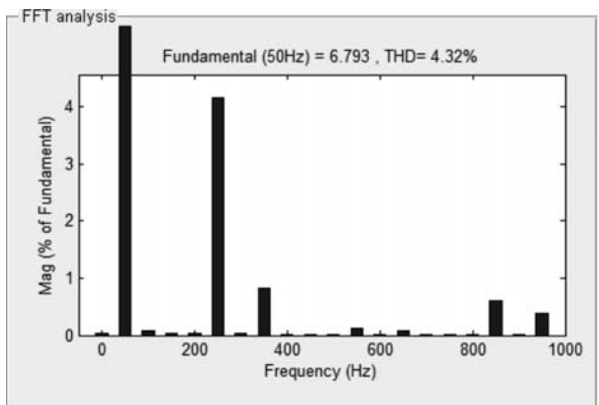

*FIG. 11. SUPPLY VOLTAGE AND CURRENT ARE IN PHASE WHEN FILTERS ARE CONNECTED*

## ACKNOWLEDGEMENTS

Authos acknowledges with thanks the higher authorities and Department of Electrical Engineering, Mehran University of Engineering & Technology, Jamshoro, Pakistan, for providing moral support and necessary facilities to complete this research work.